\documentstyle[twocolumn,aps]{revtex}
\input psfig.sty

\def\ltsim{\vbox {\hbox{\lower 0.6\baselineskip \hbox{$<$}} \break
		 \hbox{\lower 0.1\baselineskip \hbox{$\sim$}} } }
\def\gtsim{\vbox {\hbox{\lower 0.6\baselineskip \hbox{$>$}} \break
                 \hbox{\lower 0.1\baselineskip \hbox{$\sim$}} } }
\def\k{{\bf k}}

\def\R{{\bf R}}
\def\r{{\bf r}}

\def\vs{{\bf v}_s}
\begin{document}
\draft

\twocolumn[\hsize\textwidth\columnwidth\hsize\csname %
@twocolumnfalse\endcsname

\title{Vortex Contribution to Specific Heat of Dirty $d$-Wave Superconductors:
Breakdown of Scaling }

\author{
C. K\"ubert and P.J. Hirschfeld
}

\address{
Department of Physics, University of Florida, Gainesville, FL 32611, 
USA.\\
}

\maketitle
\begin{abstract}
We consider the problem of the vortex contribution to thermal properties of
dirty $d$-wave superconductors.  In the clean limit, 
the main contribution to the density of states in a $d$-wave 
superconductor arises from
extended quasiparticle states which may be treated semiclassically, giving rise to a
specific heat contribution $\delta C(H)\sim H^{1/2}$ .  We show that the
extended states continue to dominate the dirty limit, but lead to a $H \log H$ behavior
at the lowest fields, $H_{c1}\ltsim H\ll H_{c2}$.  This crossover may explain
recent discrepancies in specific 
heat measurements at low temperatures and fields
in the cuprate superconductors.  We discuss the range of validity of 
recent predictions of
scaling with $H^{1/2}/T$ in real samples. 

\end{abstract}
]
{\it Introduction.} 
 With the growing consensus that the symmetry of the cuprate superconductors
is $d$-wave\cite{Reviews} has come a renewed interest in the properties of the vortex
state in ``unconventional" superconductors with order parameter nodes.  Many of the
basic ideas about how this state differs from the conventional Abrikosov state in
classic superconductors were worked out already in the context of rotating
$^3He$ and heavy fermion superconductors.  Recently, however, a number of novel features
of the problem peculiar to those systems with Dirac spectrum (line nodes in 3D or point
nodes in 2D with order parameter vanishing linearly with angle on the Fermi surface)
have been pointed out.  Volovik\cite{Volovik1} showed that, in contrast to conventional
superconductors, extended quasiparticle states with momentum $\bf k$ near order parameter
nodal directions $\k_n$ 
dominate the density of states at zero energy.  
This leads to a specific heat which varies as $\delta C(H)\sim
H^{1/2}$, in contrast to classic superconductors, where localized quasiparticle states
in vortex cores lead to a scaling of $\delta C (H)\sim H$ since the number of vortices
scales proportionally to the field.   
Simon and Lee\cite{SimonLee} then showed that thermal and 
transport properties exhibit a  scaling with $H^{1/2}/T$, again arising simply
from the low-energy Dirac form of the electronic spectrum.

The predicted proportionality of the electronic specific heat 
to $\sqrt{H}$ was in fact identified
in  measurements on high quality single crystals by Moler et
al.,\cite{Moleretal}  one of the 
crucial early experiments lending credence to the
$d$-wave hypothesis.  However, the interpretation 
of the observed $\sqrt{H}$ dependence
has been questioned by 
Ramirez\cite{Ramirez}  who points out that
there are well-known cases where classic superconductors show 
a similar ``nonanalytic"
behavior sufficiently close to the lower critical field $H_{c1}$.  Furthermore,
experimental results of Fisher et al\cite{Fisheretal}  and Revaz et 
al.\cite{JunodLT}
cannot be well fit
by a  $\sqrt{H}$ form. 

The above scaling predictions hold, strictly 
speaking, for clean $d$-wave superconductors
and for energy scales small compared to the maximum 
gap scale $\Delta_0$.  To make
realistic predictions for experiments, deviations from scaling due to 
disorder and other
real-materials effects must be accounted for.  
In this work, we study the effects of
disorder, primarily in the unitarity scattering limit thought 
to be relevant to the
cuprates,\cite{Hotta,felds} and 
ask how the scaling predictions of refs. \cite{Volovik1,SimonLee} 
break down.
The treatment is similar in spirit to the crossover 
phenomena studied in the context
of the nonlinear Meissner effect in $d$-wave 
superconductors by Yip and Sauls\cite{YipSauls}
In the present work we are concerned with  
fields $H\gtsim H_{c1}$, however, and discuss, in a
crude way, the influence of the structure of 
the vortex state itself on thermodynamic
bulk measurements.  We analyze existing experiments and point 
out how they may be
reconciled with the $d$-wave hypothesis and 
the ideas of Volovik by properly accounting
for the effects of disorder.

{\it Treatment of extended quasiparticle states.}  
 In an external field, one should in principle solve the Bogoliubov-de
Gennes equations or equivalent for the fully self-consistent, 
spatially dependent
mean fields
and quasiparticle amplitudes.  Imposing a vortex-type boundary
condition on the order parameter around an isolated singlularity 
leads to a uniform
phase winding with corresponding superfluid velocity 
$\vs= (\hbar/2mr){\hat \theta}$,
with
$\theta$ the  azimuthal angle in real space.  The order
parameter  magnitude $\Delta_k(\R )$ is supressed near the vortex core over a
length scale of order the coherence length $\xi_0$, and has a fourfold
 symmetry in real space.\cite{1vortex}  The quasiparticle 
wavefunctions and quasiparticle
density of states have  also
been  claimed 
\cite{Makivortex} to display fourfold symmetry.  
These
excitations are of two types: bound states 
localized in the vortex cores, and
extended states which evolve smoothly into the bulk zero-field 
quasiparticle states
at large distances from the vortex.  In a 
classic superconductor, the low-temperature entropy
is dominated by the core bound states, since 
the extended states are fully gapped and therefore
essentially depopulated.  The core levels are furthermore separated by a typical 
spacing $\Delta_0^2/E_F$, where $\Delta_0$ is the gap maximum and $E_F$ is the Fermi energy.
This can be rather large in short-coherence length superconductors like the cuprates,
such that only one or a few states may actually be bound.  Even if treated as a quasicontinuum,
in a d-wave superconductor the bound states may be shown to contribute less to the
entropy than the extended states with momenta near the bulk gap nodal directions,\cite{Volovik1}
by an amount which diverges logarithmically with the vortex size or intervortex separation.
Since nonmagnetic disorder is pairbreaking in a d-wave superconductor,
the contribution of low-energy extended states to the entropy will be still greater in
the dirty systems we consider.

For the above reasons, it appears in the $d$-wave case to be a good approximation 
to ignore the  core excitations in the calculation of bulk properties.  The extended
excitations are treated here by a 
method originally proposed by Maki and Tsuneto\cite{MakiTsuneto} for classic
superconductors and  applied to the d-wave nonlinear Meissner effect by Yip and
Sauls.\cite{YipSauls}  The semiclassical approximation treats the quasiparticle states
as plane waves of energy Doppler shifted 
by $\omega\rightarrow\omega-\vs\cdot\k$.  
The single-particle matrix Green's function
for the pure system is therefore
\begin{eqnarray}
g^{(0)}(\k,\omega; \vs) = \frac
{  ( \omega-\vs\cdot\k) \tau_0 +  \Delta_{\k}\tau_1 
                         +  \xi_{\k} \tau_3 }
{(\omega-\vs\cdot\k)^2 -  \Delta_\k^2 -\xi_\k^2} \; ,
\end{eqnarray} where the $\tau_i$ are  Pauli matrices in particle-hole space.
In (1),  $\xi_k$ is the usual single-particle band measured relative to
the
Fermi level, 
and $\Delta_\k=\Delta_0\cos 2\phi$ is the bulk $d_{x^2-y^2}$ order parameter
over a cylindrical Fermi surface, taken independent of
position in real space.  Thus, we neglect both the order parameter 
suppression and the spatial variation of the
quasiparticle amplitudes near the vortex core over the coherence length
$\xi_0=\hbar v_F/\pi\Delta_0$.  
This is justified provided we confine our interest to
fields $H$  such that $H_{c1}\ltsim H\ll H_{c2}$: the coherence
length
 $\xi_0$ will then be much smaller than the other relevant length scales, 
specifically $R$, the intervortex distance and $\lambda$, the
penetration depth, in a strongly type-II system.

Disorder is introduced in the self-consistent t-matrix 
approximation\cite{HVW,SMV} through the averaged self-energy
$\Sigma_0(\omega)=\Gamma G_0 \tau_0/( c^2-G_0^2 )
$,
where $\Gamma=n_i/\pi N_0$ is an impurity scattering rate proportional
to  the concentration $n_i$ of point potential scatterers, 
$c=\cot \delta_0$ is the cotangent of the s-wave scattering phase shift
$\delta_0$, and $N_0$ is the density of states at the Fermi level.
  We focus here  on the strong scattering limit 
$\delta_0\simeq \pi/2$, but results are easily obtained for general
phase shifts.
The averaged integrated Green's function is 
$G_0(\omega) = (\pi N_0)^{-1}\Sigma_{\k} {1\over 2}\;{\rm Tr}\;\tau_0\;
g(\k,\omega) $, which for a simple $d_{x^2-y^2}$ state leads to 
$G_0(\omega)=-i(2/\pi) {\bf K}(\Delta_0/\tilde\omega)$.
This form of
the self-energy leads via the Dyson equation to an averaged propagator 
$g(\k,\omega)$ identical in form to (1)
but with $\omega$ replaced by $\tilde \omega = \omega-\Sigma_0 (\omega)$.  
Note the
Green's function must be determined self-consistently, and depends
on the single energy variable $\tilde\omega -\vs\cdot\k$.  It is
furthermore important   to observe that  
the most general form of the propagator would
include renormalizations of both the single-particle energy $\xi_\k$ and
the order parameter $\Delta_k$ as well.   Corrections due to the former vanish
identically for one-particle properties like the density of states
if the system is  particle-hole  symmetric.\cite{HWE}   Corrections due
to the latter, which vanish for a $d_{x^2-y^2}$ order parameter in the
$H=0$ case, are nonzero in general in finite field, due to the
dependence
of the shift $\vs\cdot\k$ on the angle $\phi$ over the Fermi surface.
However, for low energies such that only quasiparticles in the
neighborhood of the nodes are relevant, this renormalization may
be shown to be small,
and we have neglected it here.

{\it Density of states at zero energy in nonzero field.}
We first present results for the magnetic field dependence
of the density of states at zero energy, $N(0;H)$.
This is an experimentally accessible quantity, as it scales with the
linear-T term in the low-temperature specific heat in the superconducting state,
$\gamma_{el}=\pi^2N(0)/3$.  The origin of this term has been 
controversial, and may result in part
from nonelectronic two-level systems away from the 
planes.  The zero-field residual density of states used
in this work arises solely through disorder in the 
electronic system, and must be assumed
to represent a lower bound to the true residual 
density of states.  To find $N(0;H)$, we average
the propagator over a vortex unit cell, $N(0;H)/N_0\equiv \langle- {\rm Im}
G_0(\omega;\vs)\rangle_H$, where for any $f(\vs)$ we define
$\langle  f(\vs) \rangle_H\equiv
A^{-1}(H)\int_{\rm cell} d^2r\;  f(\vs ).
$
For simplicity, we take the cell to be 
circular, of area $A\simeq\pi R^2$. Here
$R=\xi_0(\pi/2)^{1/2}a^{-1}(H_{c2}/H)^{1/2}$ is 
the intervortex spacing, and $a$ is 
a constant of order
unity dependent only on the vortex lattice geometry.  Note that increasing
the magnetic field does not affect $\vs$, but merely decreases
   $R$.
Effects of the
actual spatial dependence of $\vs$ in the lattice phase
can be easily incorporated in the 
theory.
\indent\indent In the clean limit and for $H_{c1}\ltsim H \ll
H_{c2}$, we obtain $N(0;H)/N_0\simeq \sqrt{8/\pi}a(H/H_{c2})^{1/2}$.  
This is essentially the
result obtained by Volovik,\cite{Volovik1} who linearized the gap 
and the electronic spectrum
around the nodes, and may be derived easily by recalling that for 
the $d_{x^2-y^2}$ state
$N(\omega)/N_0\simeq \omega/\Delta_0$ for $\omega \ll 
\Delta_0$, replacing $\omega$ by $\vs\cdot\k$, and
performing the spatial integral over the cell.  
This approximation must therefore
fail at low energies, where the density 
of states in the disordered $H=0$
system has the
form $N(\omega)\sim N(0)+b\omega^2$, with $N(0)$ related to the
zero-energy quasiparticle scattering rate $2\gamma_0$ by
$N(0)=(2\gamma_0/\pi\Delta_0)\log (4\Delta_0/\gamma_0)$.  In the
unitarity limit, $\gamma_0$ is well approximated by $\gamma_0\simeq
0.61\sqrt{\Gamma\Delta_0 }$ for small concentrations.  

As in the zero-field
case above, the residual density of states depends on the quasiparticle
lifetime, which becomes, however, a local quantity in the presence of
the superflow field $\vs(\r)$.  At low energies, substitution of the
form $\tilde\omega(\r)=\omega+i\gamma(\r)$ into $\Sigma_0(\omega)$ yields for
$v_s k_F,\gamma\ll \Delta_0$ but arbitrary $v_s k_F/\gamma$,
\begin{eqnarray}
\frac{\gamma} {\Delta_0} &=& \frac{\pi}{2}\:\frac{\Gamma}
{\Delta_0}\:
\left[\right. \ln \left(
\frac{4\Delta_0}{\sqrt{\gamma^2 + (\vs\cdot \k_n)^2}}
\right)\nonumber\\
 &+& \frac{\vs\cdot \k_n}{\gamma}
        \tan^{-1} \left( \frac{\vs\cdot \k_n}{\gamma} \right)
\left.\right]^{-1}    \; ,
\end {eqnarray}
where $\k_n$ is the nodal direction.
In the ``dirty limit",  $(H/H_{c2})^{1/2}\Delta_0\ll \gamma_0 \ll\Delta_0$,
the spatial integrations in $\langle -{\rm Im }G_0 \rangle_H$
can be performed, yielding 
\begin{equation}
{\delta N(0,H)\over N_0}\simeq {\Delta_0\over 8\gamma_0}
a^2\left ({H\over H_{c2} } \right) \log  
\left[ {\pi\over 2a^{2}}\left ({H_{c2}\over H}  \right)  \right]  
\end{equation}
%
%
\begin{figure}[h]
\leavevmode\centering\psfig{file=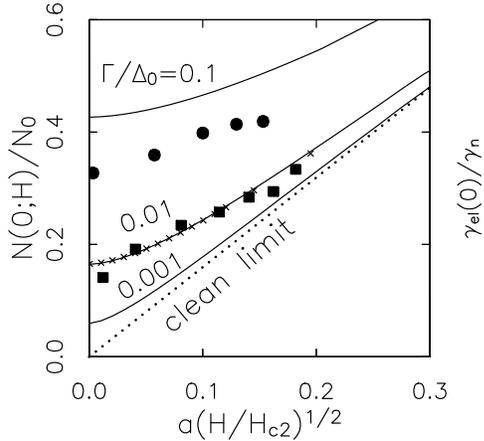,width=2.6 in}
\caption{Normalized density of states $N(0;H)/N_0$ for
$\Gamma/\Delta_0=0.1$, 0.01, and 0.001 in unitarity limit $c=0$.
Nodal approximation (solid lines), exact result (crosses), clean, low
energy approximation (dotted line).
Data 
from Fisher et al.\protect\cite{Fisheretal} (circles);
Moler et al.\protect\cite{Moleretal} (untwinned sample, squares), 
assuming $H_{c2}/a^2$=300T,
$\gamma_n=15$ mJ-mol-${\rm K}^2$. }
\end{figure}
To illustrate the deviations from the clean limit, low-
energy result,
we
plot in Figure 1 a full numerical
evaluation of the  density of states N(0;H).
For the cleanest case plotted, the predicted scaling is attained over
a substantial range.  Deviations from the  $\sqrt{H}$ behavior occur
below a lower crossover field scale $H^*\equiv (\gamma_0/a\Delta_0)^2
H_{c2}$, corresponding to the smearing of the linear density of states
due to  scattering by impurities. 
  We note that excellent approximate results  may be
obtained,
with a substantial increase in speed of numerical evaluation, by replacing
$\vs\cdot\k$  in $g(\k,\omega;\vs)$ by its value near the nodes,
$\vs \cdot \k_n$, so that $G_0(\omega;\vs)\simeq 
(1/\pi)\sum_{\alpha=\pm}{\bf K}(\Delta_0/(\tilde\omega-\alpha\vs\cdot\k_n))$.
  We use this approximate
form, whose validity is shown in  Figure 1, in all further calculations.

In Figure 1 we have also shown experimental 
data of Moler et al.\cite{Moleretal} and
Fisher et al.\cite{Fisheretal} We have plotted published data for the linear
electronic specific heat coefficient
$\gamma_{el}$ without attempting to subtract residual ``extrinsic" contributions
possibly due to 2-level systems, etc.   
We observe that the data for the untwinned 
YBCO crystal of Moler et al, which are claimed to follow the clean-limit
prediction of Volovik,\cite{Volovik1} are actually consistent with a slightly
dirty $d$-wave superconductor as well if the entire residual contribution
to $N(0;H)$ is attributed to impurities (analysis of NMR 
and penetration depth data on similar samples makes
this interpretation unlikely, however.\cite{Moleretal}).  No good fit
to the clean limit prediction was obtained for the sample studied by Fisher
et al.  From the Figure it is clear that this discrepancy might be explained
by simply assuming the Fisher et al. sample contains roughly an order of magnitude
more defects than the Moler et al. sample.  Clearly  systematic
disorder studies would be very useful in resolving these questions.

{\it Density of states at finite frequency.} Kopnin and 
Volovik\cite{KopninVolovik1} have
observed that the field-dependent part of the density of states, 
$\delta N(\omega;H)$  
will diverge at low frequencies as $1/\omega$
down to a lower cutoff of order the average quasiparticle energy shift
$E_H\equiv a(H/H_{c2})^{1/2}\Delta_0$.
In the current formalism, this result is recovered by noting that   
the field-dependent part of the local
density of states is given by $\langle -{\rm Im}[G_0(\omega;\vs)-G_0(\omega)]
\rangle_H$, which for
$ \gamma_0<E_H , \omega\ll \Delta_0$ yields 
\begin{eqnarray}
&&{\delta N(\omega;H)\over N_0}\simeq \langle\left[
\left|{\vs\cdot\k_n-\omega\over \Delta_0}\right| - \left|{\omega\over
\Delta_0}\right|\right] 
\rangle_H
\simeq {E_H\over \sqrt{2\pi\Delta_0^2}}F(x)\\&&F(x)\equiv{1\over x}\left\{ 
\begin{array}{ll}
\pi/ 2 & x >1\\
3x\sqrt{1-x^2}+(1+2x^2)\sin^{-1}x-\pi x^2 ~&  x<1
\end{array}\right.~~~\nonumber
\end{eqnarray}
where $x=\sqrt{2/\pi}(\omega/ E_H)$.  The desired result is then 

\begin{figure}[h]
\leavevmode\centering\psfig{file=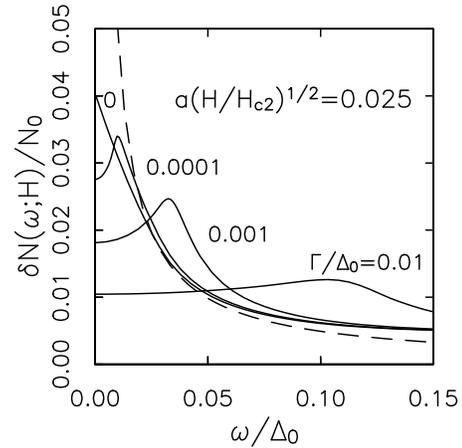,width=2.6 in}
\caption{Density of states at $a(H/H_{c2})^{1/2}=0.025$ vs. 
freq. $\omega/\Delta_0$ for $\Gamma/T_c$=0.,0.0001,0.001, 0.01 (solid
lines).  Intermediate-frequency asymptotic result $\delta N \simeq
a^2\pi \Delta_0 H/(4 \omega H_{c2})$ (dashed line).}
\end{figure}
\noindent obtained in the limit
of large frequencies,  
$\delta N(\omega;H)/N_0\simeq a^2\pi\Delta_0H/(4\omega
H_{c2})$ .
 At energies smaller than the
average quasiparticle energy shift $E_H$, this divergence is cut off
as 
$\delta N(\omega;H)\simeq
N(0;H)(1-\pi x/4))$.  Both these limits are clearly visible in the clean
case shown in Figure 2.
The $1/\omega$ divergence may also be cut off in a different fashion if
the impurity scale $\gamma_0$ exceeds the magnetic 
energy,
$E_H<\omega<\gamma_0$, as also shown (the maximum in each curve corresponds
roughly to the scale $\gamma_0$).  In dirty samples, such that
$\gamma_0$ becomes an appreciable fraction of the gap scale, the $1/
\omega$ behavior
will be unobservable.

{\it $H^{1/2}/T$ scaling of specific heat}.  The specific heat is now easy
to calculate by differentiating the entropy of a free Fermi gas of
quasiparticles with disorder- and field-averaged density of states
$ N(\omega;H) $; one finds at low temperatures:
\begin{eqnarray}
C&\simeq&2\int_0^\infty d\omega \left(\omega\over T\right)^2
\left({-\partial f\over \partial \omega} \right)  N(\omega;H)
\\
&\simeq&\left\{ \begin{array}{ll}  N(0;H) {\pi^2\over 3} T        
~~ &T \ll {\rm max}[\gamma_0,E_H]\ll \Delta_0 \\
N_0 \left(9\zeta(3) \over \Delta_0 \right)T^2~~~~ & \gamma_0,E_H\ll T \ll \Delta_0
\end{array}\right.\nonumber
\end{eqnarray}
A $T^2$ term  characteristic of the pure $d$-wave system in zero field is
present whenever both the impurity and magnetic field scales are smaller
than the temperature.   If one 
\begin{figure}[h]
\leavevmode\centering\psfig{file=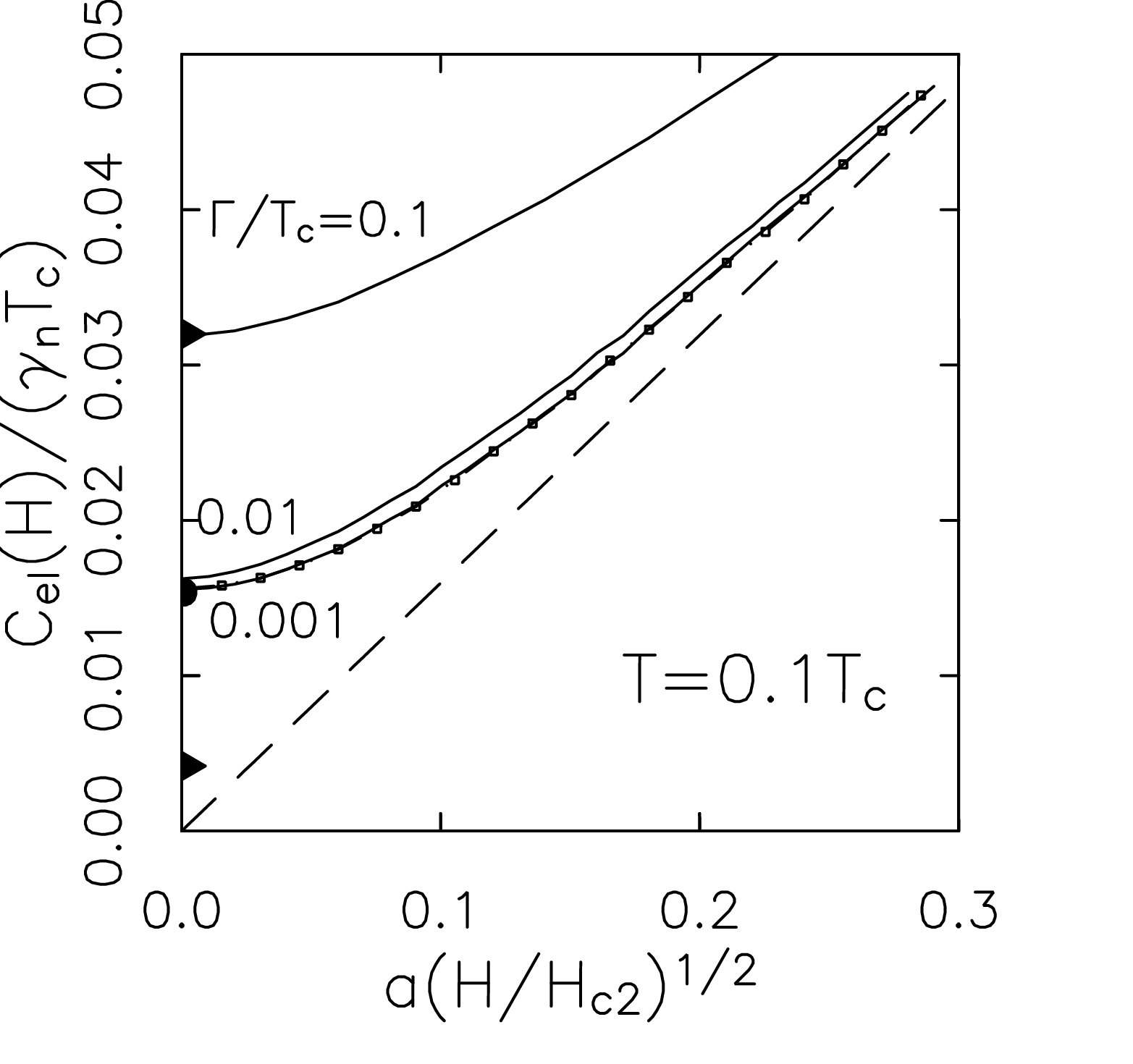,width=2.6 in}
\caption{Electronic specific heat $C_{el}(H)/(\gamma_{n}T_c)$  at $T=0.1T_c$ 
normalized to normal state value, $\gamma_n T_c$ vs. square root
of field, $a(H/H_{c2})^{1/2}$ for $\Gamma/T_c=$ 0.1, 0.01 and 0.001
(solid lines).  Clean limit (small filled circles).
Asymptotic low-$T$ clean limit (dashed line).  Zero-field linear
term $N(0;0)T$ (filled triangles) for values of $\Gamma$ shown.}  
\end{figure}
\noindent plots the specific heat at fixed temperature,
as in Fig. 3, this results at low fields in deviations from $\sqrt{H}$ 
behavior due both to impurity
induced residual density of states and, in cleaner samples,  the $T^2$ term.
The latter effect is the origin of the saturation of the progressively cleaner
curves in the figure to a non-$\sqrt{H}$ behavior.

Finally, we examine the scaling behavior predicted for
the
specific heat.\cite{SimonLee,KopninVolovik2}  
In the clean, low-energy limit, 
\begin{figure}[h]
\leavevmode\centering\psfig{file=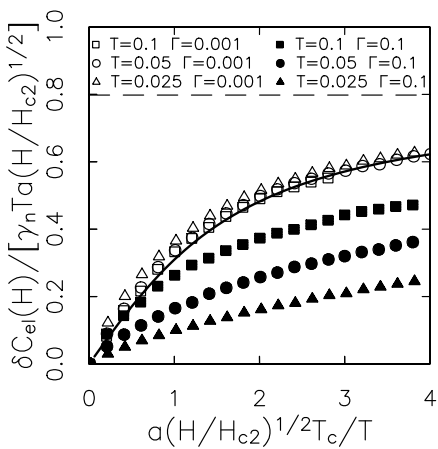,width=2.6 in}
\caption{Normalized vortex contribution to specific heat, 
$\delta C_{el}(H)/[\gamma_{el}Ta(H/H_{c2})^{1/2}]$ vs.
$Y\equiv a(H/H_{c2})^{1/2}T_c/T$ for fixed temperatures $T$ and
scattering rates $\Gamma$ as shown; unit of energy $T_{c0}$.
Asymptotic large-$Y$ limit $\sqrt{2/\pi}$ (dashed line).
}
\end{figure}
\noindent  this result may be
derived by substituting Eq. (4) into (5).  
One finds $\delta C(H)/[\gamma_nTa(H/H_{c2})^{1/2}]=F_C[Y]$,
where $Y=a(H/H_{c2})^{1/2}T_c/T$ and $F_C$ is a 
scaling function which varies as 
$F_C\simeq 3\log 2 \Delta_0 Y/(4\pi T_c)$ for $Y\ll 1$ and as 
$F_C\simeq \sqrt{2/\pi}$ for
$Y\gg 1$.  The full numerically determined, 
low-energy clean limit scaling function is plotted
in Figure 4, along with full numerical evaluations 
of (5), which  includes corrections beyond
the low-energy approximation, as well as 
impurity effects.  Scaling is expected for a given data set 
provided $H,T$ are such that $E_H$ and $T$ 
are both larger than the impurity scale $\gamma$. 
For example, in the clean case chosen (open symbols), scaling
is 
obtained over the full range of $Y$, whereas for the dirty 
system (filled symbols) scaling has broken down completely.
Junod\cite{Junodscaling} has recently reported data which crudely
follow scaling predictions, but scatter in the data is too great
to ascertain if the deviations follow the pattern predicted here.

{\it Conclusions.}
We have placed the elegant scaling arguments 
of Volovik\cite{Volovik1,KopninVolovik2} and Simon and Lee\cite{SimonLee}
regarding the 
 specific heat of a $d$-wave
superconductor in magnetic field 
on  a concrete foundation, introducing a simple formalism 
capable of including both the effects of disorder and of energies
comparable to the gap scale. We have shown that
 the density of
states of a dirty $d$-wave
system varies as $H \log H$ rather than the $\sqrt{H}$ expected
for the clean system.  As a result, the predicted scaling of
the vortex specific heat $\delta C(H)$ with $\sqrt{H}/T$
breaks down in a well-defined fashion.  As sample variability in the cuprates
is notorious, it will be important to understand these deviations.  
We are in the process of studying similar effects in transport 
properties in a magnetic field.

{\it Note added:} After submission of this paper
we learned that Y. Barash et al.\cite{Barash}  
had also obtained Eq. (3).  We are
grateful to G.E. Volovik for pointing out this reference.

{\it Acknowledgments.}  The authors gratefully acknowledge 
discussions with A. Dorsey and P. W\"olfle.  
Partial support was provided by
NSF-DMR-96--00105 and by the A. v. Humboldt Foundation (CK).

\end{document}